\documentclass[structabstract]{aa}  

\usepackage{graphicx}
\usepackage{txfonts}
\usepackage[percent]{overpic}
\usepackage{color}
\usepackage{multirow}
\usepackage{natbib,twoopt} 
\usepackage[breaklinks=true]{hyperref} 
\bibpunct{(}{)}{;}{a}{}{,} 
\begin{document}

   \title{Signatures of running penumbral waves in sunspot photospheres}

   \author{J. L\"ohner-B\"ottcher \and N. Bello Gonz\'alez }
    \institute{Kiepenheuer-Institut f\"ur Sonnenphysik, Sch\"oneckstr. 6, 79104 Freiburg\\
              \email{jlb, nbello @kis.uni-freiburg.de}}
   \date{Received 31 March 2015 / Accepted 19 June 2015}

  \abstract
  {The highly dynamic atmosphere above sunspots exhibits a wealth of magnetohydrodynamic (MHD) waves. Recent studies suggest a coupled nature of the most prominent phenomena: umbral flashes and running penumbral waves (RPWs).}
  {From an observational point of view, we perform a height-dependent study of RPWs, compare their wave characteristics, and aim to track down these so far only chromospherically observed phenomena to photospheric layers to prove the upward propagating field-guided nature of RPWs.}
  {We analyze a time series (58\,min) of multiwavelength observations of an isolated circular sunspot (NOAA11823) taken at high spatial and temporal resolution in spectroscopic mode with the Interferometric BIdimensional Spectro-polarimeter (IBIS/DST). By means of a multilayer intensity sampling, velocity comparisons, wavelet power analysis, and sectorial studies of  time slices, we retrieve the power distribution, characteristic periodicities, and propagation characteristics of sunspot waves at photospheric and chromospheric levels.}
  {Signatures of RPWs are found at photospheric layers. Those continuous oscillations occur preferably at periods between \mbox{4-6\,min} starting at the inner penumbral boundary. The photospheric oscillations all have a slightly delayed, more defined chromospheric counterpart with larger relative velocities, which are linked to preceding umbral flash events. In all of the layers, the power of RPWs follows a filamentary fine-structure and shows a typical ring-shaped power distribution increasing in radius for larger wave periods. The analysis of time slices reveals apparent horizontal velocities for RPWs at photospheric layers of $\approx51\,\rm{km/s,}$ which decrease to $\approx37\,\rm{km/s}$ at chromospheric heights. The photospheric distribution of peak periods at the inner penumbra resembles the chromospheric cases.}
  {The observations strongly support the scenario of RPWs being upward propagating slow-mode waves guided by the magnetic field lines. Clear evidence for RPWs at photospheric layers is given. Assuming an inverse proportionality of the peak period and cut-off period on the cosine of the field inclination, the penumbral magnetic field inclination is increasing toward the outer penumbra. The more rapid increase and the larger horizontal velocities at photospheric heights hint at the more horizontal penumbral field inclination at lower heights.}

  \maketitle
  \titlerunning{Signatures of running penumbral waves in the sunspot photosphere} 
  \authorrunning{L\"ohner-B\"ottcher \& Bello Gonz\'alez}

\section{Introduction}\label{sec_intro}
Sunspot waves are one of the most spectacular and dynamic phenomena in the solar atmosphere. Since their first detection, the question about the nature of umbral flashes \citep[]{1969SoPh....7..351B} and running penumbral waves \citep[][henceforth RPWs]{1972SoPh...27...71G,1972ApJ...178L..85Z} and their relation has been under intense discussion. Whereas umbral flashes have been commonly interpreted as upward propagating magnetoacoustic slow-mode waves guided by magnetic field lines \citep[e.g., ][]{2006ApJ...640.1153C,2013A&A...556A.115D}, the case for RPWs is not yet clear. 

A first scenario suggests that RPWs are purely chromospheric waves excited by umbral flashes and propagate radially outward across the chromospheric penumbra \citep{2006A&A...456..689T}. Secondly, recent studies support the scenario that running penumbral waves,  like umbral flashes,  are channeled, upward propagating waves excited at lower layers \citep[e.g.,]{2003A&A...403..277R,2006SoPh..238..231K,2007ApJ...671.1005B,2013ApJ...779..168J}. The increasing magnetic field inclination toward the outer penumbra would explain this visual pattern of horizontal propagation at a chromospheric layer since the increasing path length for a coherent wave front delays its occurrence. However, a detailed photospheric observation focusing on RPWs to disprove the purely chromospheric nature of RPWs was still lacking until now.

Under the assumption of field-guided waves, the peak period or frequency in a power spectrum of sunspot waves can be used as an indicator for the magnetic field inclination. The acoustic cut-off, which reduces waves with periods above a certain cut-off value, is highly dependent on the field topology \citep{1977A&A....55..239B}. The effective cut-off period increases with the inclination angle of the field lines. Observational studies \citep[e.g.,][]{2013ApJ...779..168J,2014A&A...561A..19Y} show that the peak period in chromospheric power spectra increases from $3\,\rm{min}$ in the umbral area to around $10\,\rm{min}$ toward the outer penumbra. For further information, see \citet{2006RSPTA.364..313B}.

\begin{figure*}[htbp]
\begin{center}
\includegraphics[trim = 0cm 0cm 0cm 0.9cm,clip,width=18.5cm]{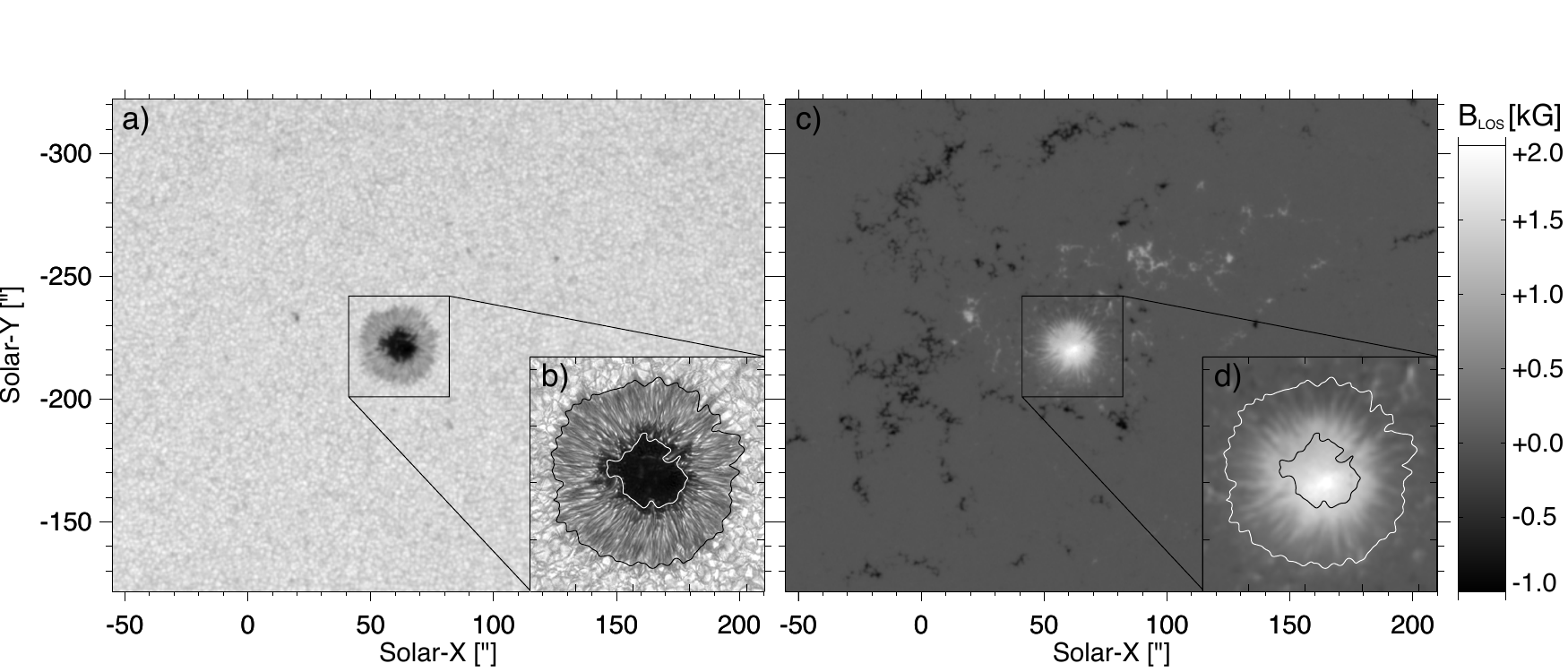}
\caption{NOAA11823 as seen in continuum (left panels) and line-of-sight magnetic field strength (right panels) on August 21st 2013 at 15:00:45UTC as observed by HMI (a+c+d) and ROSA (b). The black squares mark the analyzed region. The insets (b+d) show the sunspot with contours indicating the inner and outer penumbral boundary from the time-averaged HMI continuum intensity. The HMI magnetogram is scaled from $-1\,\rm{kG}$ (black) to $2\,\rm{kG}$ (white).}
\label{fig5}
\end{center}
\end{figure*}

Most studies on RPWs have focused on the chromosphere. Only a few attempts \citep[e.g.,][]{1998ApJ...497..464L,2007ApJ...671.1005B} have been made to determine their photospheric behavior in detail. In this work, we perform a novel high-resolution, multilayer study concentrating on the photospheric characteristics of RPWs. In Sect.\,\ref{sec_data} of this article, we give a brief description of the observations, the data, and calibration techniques. In Sect.\,\ref{sec_results}, we present the wave analysis and discuss the results. Finally, in Sect.\,\ref{sec_conclusions} we draw our conclusions and describe the importance of the new findings.

\section{Observations and data analysis}\label{sec_data}
\begin{figure*}[htbp]
\begin{center}
\includegraphics[width=17.5cm]{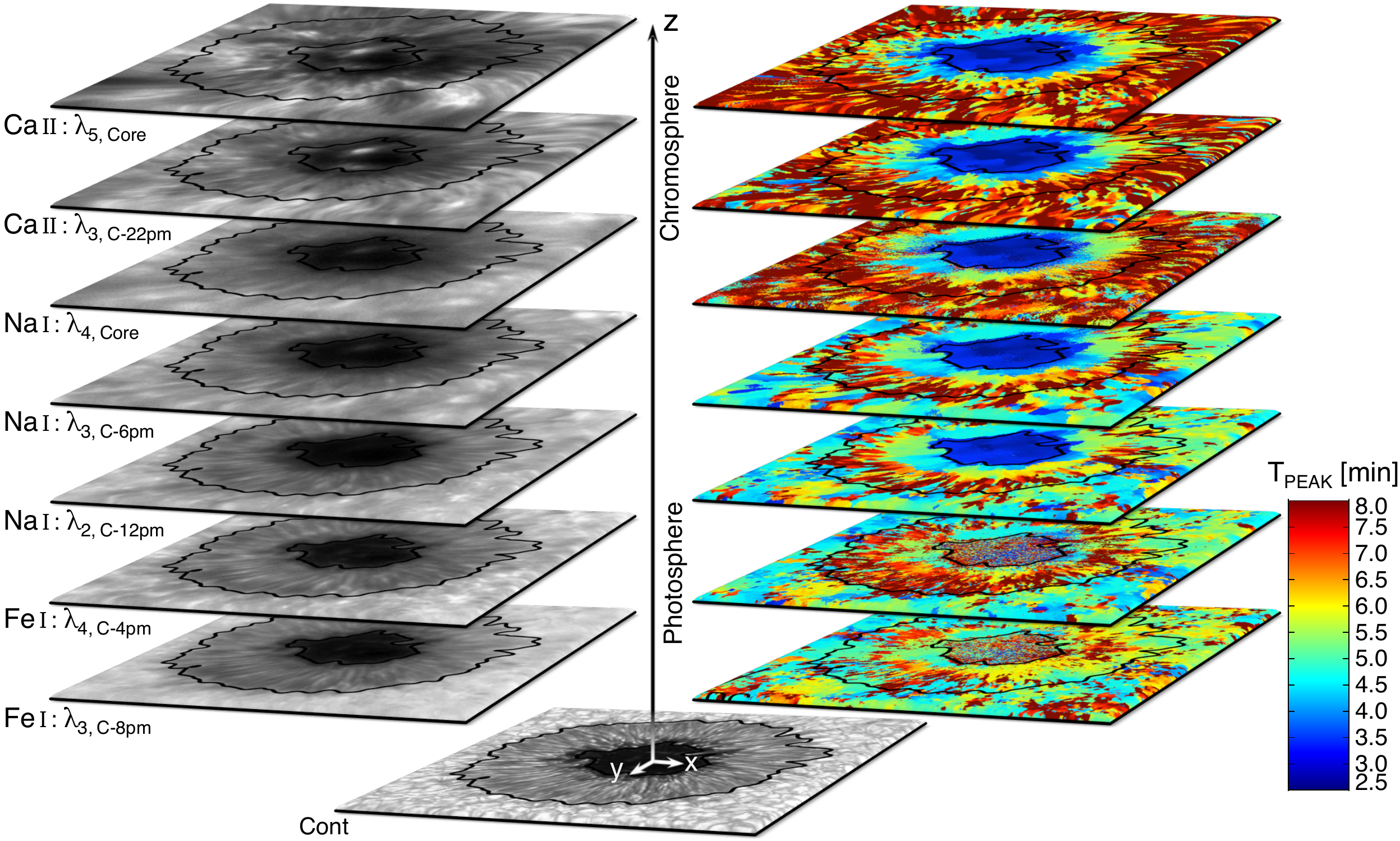}
\caption{Three-dimensional view of intensities and peak periods of intensity wave power of NOAA11823 at various wavelength positions. The intensities (left) show the sunspot at August 21st 2013 at 15:00:06UTC ($\pm3{\rm\,s}$). The images along the z-axis belong to several line core and wing positions of \ion{Fe}{I} $630.15\,\rm{nm}$, \ion{Na}{I} $589.6\,\rm{nm,}$ and \ion{Ca}{II} $854.2\,\rm{nm}$. The corresponding time-averaged (\mbox{$\approx1\rm{h}$}) distribution of peak periods $T_{\rm PEAK}$ of the intensity wave power is shown on the right.
The periods are scaled from 2.5\,min (dark blue) to 8\,min (dark red). The black contours indicate the location of the umbra (inner) and penumbra (outer) in continuum intensity (bottom panel). Whereas the length of the axis arrows represent distances around 1.5\,Mm, the image positions along the z-axis are not to scale.}
\label{fig1}
\end{center}
\end{figure*}

To study the wave phenomena in a sunspot (see Fig.\,\ref{fig5}) from photospheric to chromospheric layers, observations were carried out on August 21st 2013 from 14:53UTC to 15:51UTC at the Dunn Solar Telescope (DST) at the National Solar Observatory in New Mexico. The atmospheric conditions, which are crucial for ground-based observations, were excellent and stable allowing an effective spatial resolution of up to $0.4\,\arcsec$ even for the nonreconstructed data. Simultaneous multi-instrument, multiwavelength observations were performed with the etalon-based imaging spectropolarimeter IBIS \citep{2006SoPh..236..415C} in spectrometric service mode together with the broadband imaging instrument ROSA \citep{2010SoPh..261..363J} to sample the sunspot's photosphere and chromosphere (see left panels of Fig.\,\ref{fig1}). A complete and detailed overview of the data will be given in a forthcoming article, including a full time lapse movie of the observations (L\"ohner-B\"ottcher \& Bello Gonz\'alez, in preparation).

In this work, we focus on the narrowband spectrometric observations done with the Interferometric BIdimensional Spectro-polarimeter (IBIS) for the spectral lines \ion{Fe}{I} $630.15\,\rm{nm}$, \ion{Na}{I}\,D1 $589.6\,\rm{nm}$ and \ion{Ca}{II} $854.2\,\rm{nm}$. For the penumbra, the \ion{Fe}{I} $630.15\,\rm{nm}$ intensities stem from photospheric height of 0\,km (line continuum) to 300\,km (line minimum) above \mbox{$\tau_{\rm 500nm}=1$} \citep{2005A&A...434..317B}. Beyond that, the \ion{Na}{I}\,D1 $589.6\rm{nm}$ and \ion{Ca}{II} $854.2\rm{nm}$ lines cover the sunspot's atmosphere up to chromospheric layers. Each spectral line was sampled at $10-12$ nonequidistant wavelength positions (exposure time of $35\,\rm{ms}$). The lines were scanned successively, leading to a cadence of $13.2\,\rm{s}$. According to the Nyquist criterion, the oscillations with periods down to $26\,\rm{s}$ can be investigated.

The isolated, perfectly stable sunspot of NOAA11823 is shown in Fig.\,\ref{fig5} in continuum intensity (panels a+b) and line-of-sight magnetic field strength (panels c+d). The sunspot has a fully-developed circular penumbra, a diameter of 24 Mm and is located close to disk center at a heliocentric angle $\theta = 14^\circ$ at disk coordinates $(X,Y)=(63\arcsec,-222\arcsec)$. The HMI magnetograms revealed the unipolar direction of the magnetic field and a maximum umbral field strength of +2.2\,kG. As continuum intensity and magnetic field strength show, some small and stable pores surround the sunspot at a distance of more than $30\arcsec$ from the outer penumbral boundary. The reduced field of view, indicated by Figs.\,\ref{fig5}\,b+d, has a size of $42\,\rm{Mm}^2$ at a pixel scale of $0.098\,\arcsec\,\rm{px}^{-1}$. 

The data calibration of the IBIS data involved background intensity subtraction, flat-field calibration, correction for collimated wavelength shifts and prefilter transmission, as well as an iterative reduction of temporal image distortions caused by atmospheric turbulences. Doppler velocities for \ion{Fe}{I} $630.15\,\rm{nm}$ and \ion{Na}{I}\,D1 $589.6\,\rm{nm}$ were calculated as relative shifts of line minimum positions of a Gaussian approximation to the full line profile. The calculated velocities therefore represent the averaged motion over the atmospheric layers these lines form in. For the penumbra, this yields reliable results since the oscillatory line shifts dominate the effects of the magnetic field on the line profiles. The accuracy of the photospheric velocities is confirmed by the Doppler velocities from the Helioseismic and Magnetic Imager (HMI/SDO). These continuous and stable satellite data using the \ion{Fe}{I} line at $617.3\,\rm{nm}$ with a cadence of $45 \,\rm{s}$ and spatial scale of $0.5\,\arcsec\rm{px}^{-1}$ serve as context information for our observations (see Fig.\,\ref{fig3}).

For the investigation of characteristic frequencies and power distribution, we performed a time-position-dependent wave power analysis based on wavelet techniques \citep{1998BAMS...79...61T} as in \citet{2010A&A...522A..31B}. On the basis of a continuous oscillatory stability, we limit this study to the interpretation of the global (time-averaged) wavelet power spectra calculated at a fine nonlinear scale sampling. One hundred period steps sample the range from 26\,s to 14\,min; the step size increases to higher periods. The characteristic interval of sunspot waves between 2\,min and 6\,min is scanned by 34 steps. For the power analysis with respect to the distance to the sunspot center, we included spherical projection effects implied by the sunspot location and orientation. We therefore used foreshortened circles to compute the azimuthal average as in \citet{2013A&A...551A.105L}. Close to disk center, these effects are small but should not be neglected. For the temporal evolution of RPW velocities in radial direction, we also used the projection of circular sectors.

\section{Results and discussion}\label{sec_results}
The analysis of this high-resolution, multiwavelength observation of a perfectly stable and symmetrical sunspot yields clear evidence for photospheric RPWs and their field-guided propagation to higher layers. 

\begin{figure}[htbp]
\begin{center}
\includegraphics[width=9.0cm]{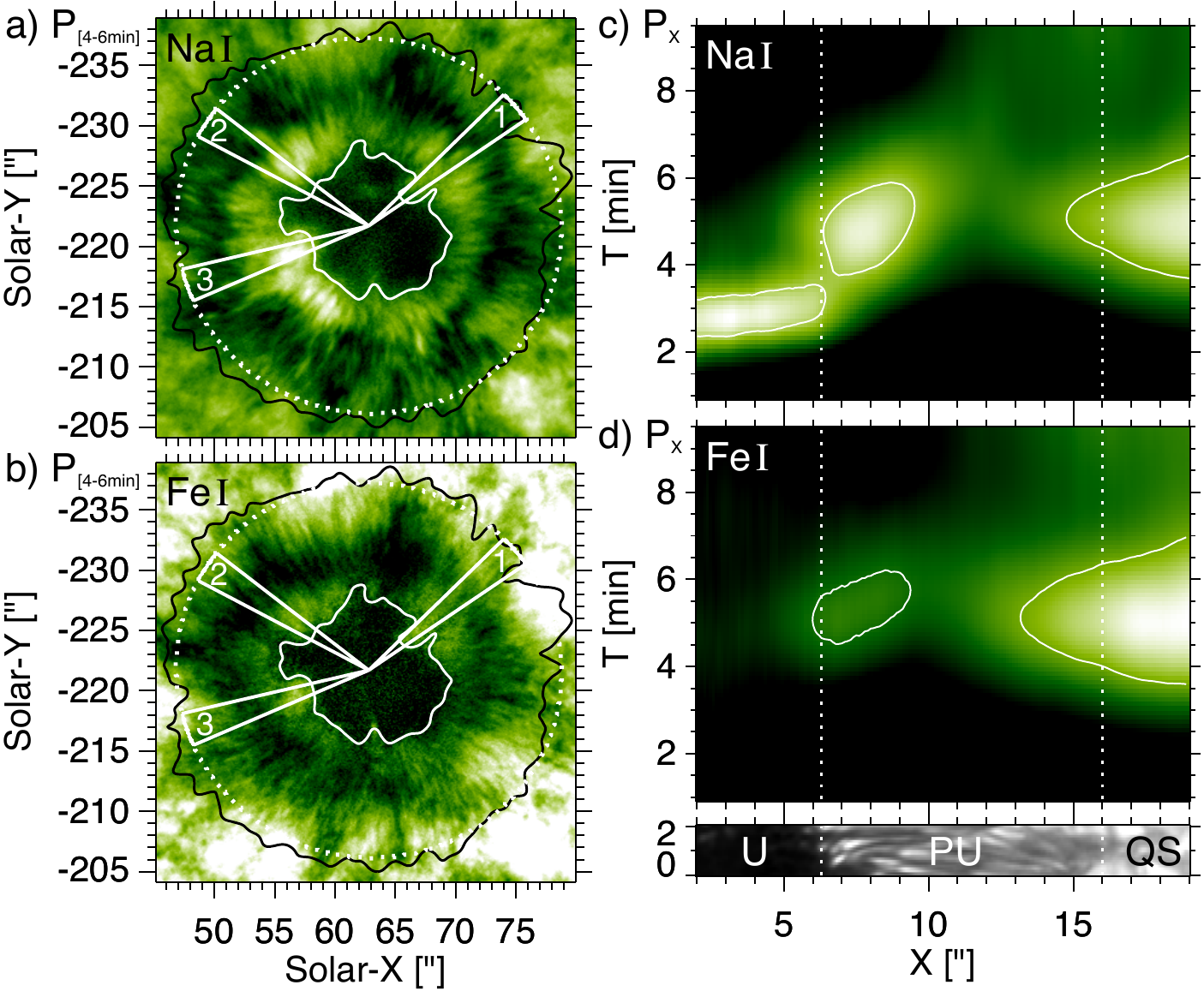}
\caption{Spatial distribution of the time-averaged wavelet power $P$ of the Doppler velocities from \ion{Na}{I} 589.6\,nm (upper panels) and \ion{Fe}{I} 630.15\,nm (lower panels). Panels a and b show the power in the 4-6\,min band. The power (bright colors indicate high power) is plotted in logarithmic (arbitrary) units to assure best contrast in the penumbra. The contours mark the inner and outer penumbral boundary in continuum intensity. Circular sectors 1-3 (white solid) follow the foreshortened projection of a circle (dotted). Panels c and d show the azimuthally averaged global power spectrum $P_{\rm X}$ for periods $T$ (in min) according to the distance X (in arcsec) from the spot center. The white contours mark the important elements in the spatial power analysis. For spatial comparison, an arbitrary radial element in continuum intensity is added below. The dashed lines indicate the inner and outer penumbral boundaries.}
\label{fig2}
\end{center}
\end{figure}

\begin{figure}[htbp]
\begin{center}
\includegraphics[width=9.0cm]{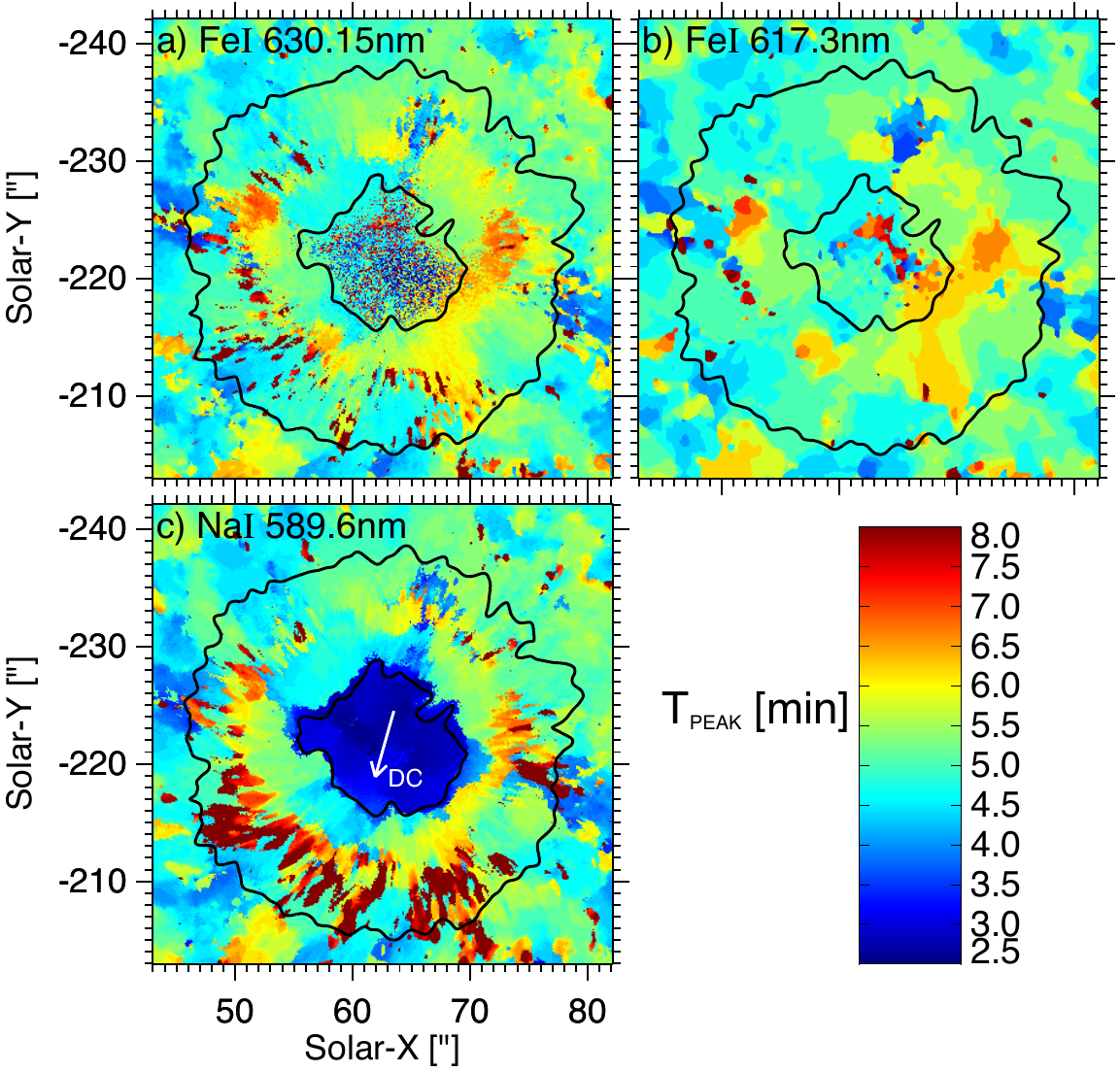}
\caption{Spatial distribution of the peak periods from the global power spectra of Doppler velocity oscillations. The major periods $T_{\rm PEAK}$ (in min) are shown for a) \ion{Fe}{I} 630.15\,nm, b) \ion{Fe}{I} 617.3\,nm (HMI), and c) \ion{Na}{I} 589.6\,nm. The scale ranges from 2.5\,min (dark blue) to 8\,min (dark red). The black contours mark the umbral and penumbral boundaries in continuum intensity. The white arrow is pointing to the disk center.}
\label{fig3}
\end{center}
\end{figure}

The wave power analysis ($P$) for Doppler velocities from the photospheric \ion{Fe}{I} $630.15\,\rm{nm}$ and photospheric-chromospheric \ion{Na}{I}\,D1 $589.6\,\rm{nm}$ lines reveals the spatial distribution and magnitude of the oscillatory features in the sunspot's atmosphere. As shown in Fig.\,\ref{fig2}, the power of the RPWs is concentrated in the inner penumbra for both layers (bright colors indicate high power). Whereas the first RPW signatures are detected at the umbral boundary at periods from 3-4\,min, the main power peaks between 4-6\,min as filamentary structures in the inner half of the penumbra. The averaged power in this period range is shown in Fig.\,\ref{fig2}\,a and b. The mentioned fine-structured pattern can also be seen in the calculated peak period maps (Figs.\,\ref{fig1} and \,\ref{fig3}), and seems to reflect the penumbral spine-intraspine distribution in intensity. 

As the illustrative contours in Fig.\,\ref{fig2}\,c and d point out, with increasing distance in the inner penumbra the RPW power $P_{\rm X}$ shifts toward larger periods. The shift in period at the inner penumbra stands in contrast to the constant {\em p}-mode signal in the 4-6\,min range at the outer penumbra and quiet sun. This evidence for RPWs is even more prominent in the low chromosphere where the shift to higher periods is still traceable to the outer penumbra up to periods of around 10\,min. To verify this radial increase of wave periods, which for chromospheric heights is in line with recent studies \citep{2013ApJ...779..168J,2014A&A...561A..19Y}, the peak periods in the power spectra of both velocity (Fig.\,\ref{fig3}) and intensity (Fig.\,\ref{fig1}) oscillations were determined for all heights. Whereas at photospheric layers the radial transition to larger peak periods (from 4\,min to 8\,min) is centered in the innermost penumbra, at higher layers the increase in period is smoother. In chromospheric heights, larger periods are therefore located further outside in the penumbra and beyond. 

The height-sampled distribution of wave power and shift in peak period hint at the opening field topology of sunspots. According to cut-off theory and observations \citep{1977A&A....55..239B,2004Natur.430..536D,2006A&A...456..689T}, the observed peak period $T_{\rm Peak}$ of field-guided waves depends on the acoustic cut-off period $T_{\rm cut,\phi}$ and increases with temperature $\vartheta_{\rm K}$ and inclination angle $\phi$ (with respect to the vertical) qualitatively like $T_{\rm Peak}\sim T_{\rm  cut,\phi}\,/1.25\sim \sqrt{\vartheta_{\rm K}}\,/\cos\phi$. Larger inclinations would especially allow the propagation of overpowering high-period waves into higher atmospheric layers. Assuming the acoustic cut-off already occurring  for photospheric RPWs, the larger periods would indicate a more horizontal orientation of the penumbra at photospheric than at chromospheric levels. In addition to the spectroscopic study of IBIS data, the analysis of HMI Doppler velocities and their power distribution confirm the photospheric signatures of RPWs, though at a lower spatial and temporal resolution (compare Fig.\,\ref{fig3}\,a+b).

\begin{figure}[htbp]
\begin{center}
\includegraphics[width=9.0cm]{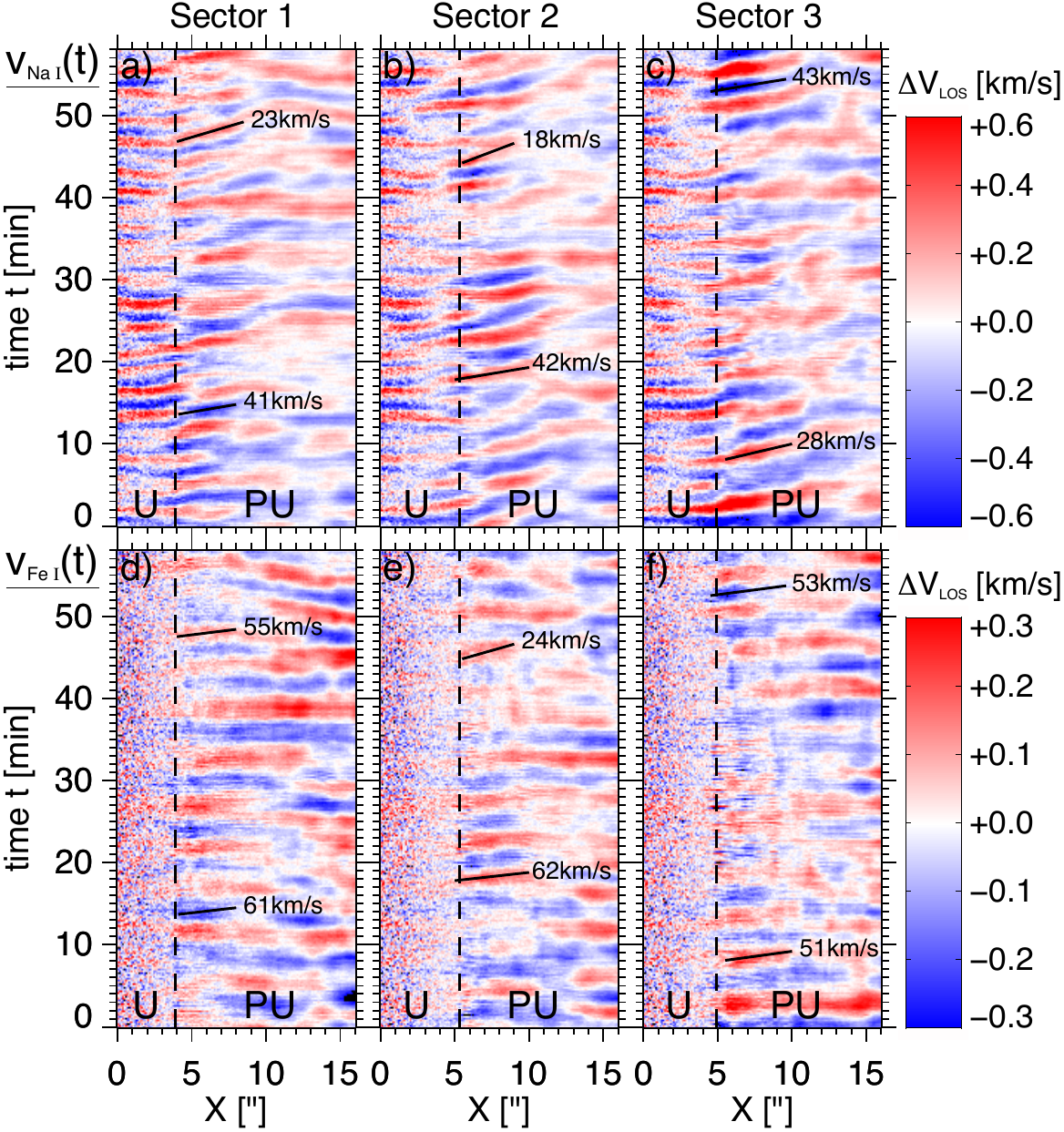}
\caption{Temporal evolution (in min) of the relative velocities in the sunspot atmosphere. The analysis of  time slices was performed for the Doppler velocities $\rm{V}_{\rm \ion{Na}{I}}$ (panels a-c) of \ion{Na}{I} 589.6\,nm and $\rm{V}_{\rm \ion{Fe}{I}}$ (d-f) of \ion{Fe}{I} 630.15\,nm in all three sectors, also shown in Fig.\,\ref{fig2}. For each sector, the azimuthal average at a radial distance X (in arcsec) from the spot center was calculated. The dashed lines mark the umbral boundary. The velocity scale $\Delta \rm{V}_{\rm LOS}$ ranges between $\pm0.6\,\rm{km/s}$ (a-c) and $\pm0.3\,\rm{km/s}$ (d-f). The black bars trace the apparent wave trains.}
\label{fig4}
\end{center}
\end{figure}

\begin{figure}[htbp]
\begin{center}
\includegraphics[width=9.0cm]{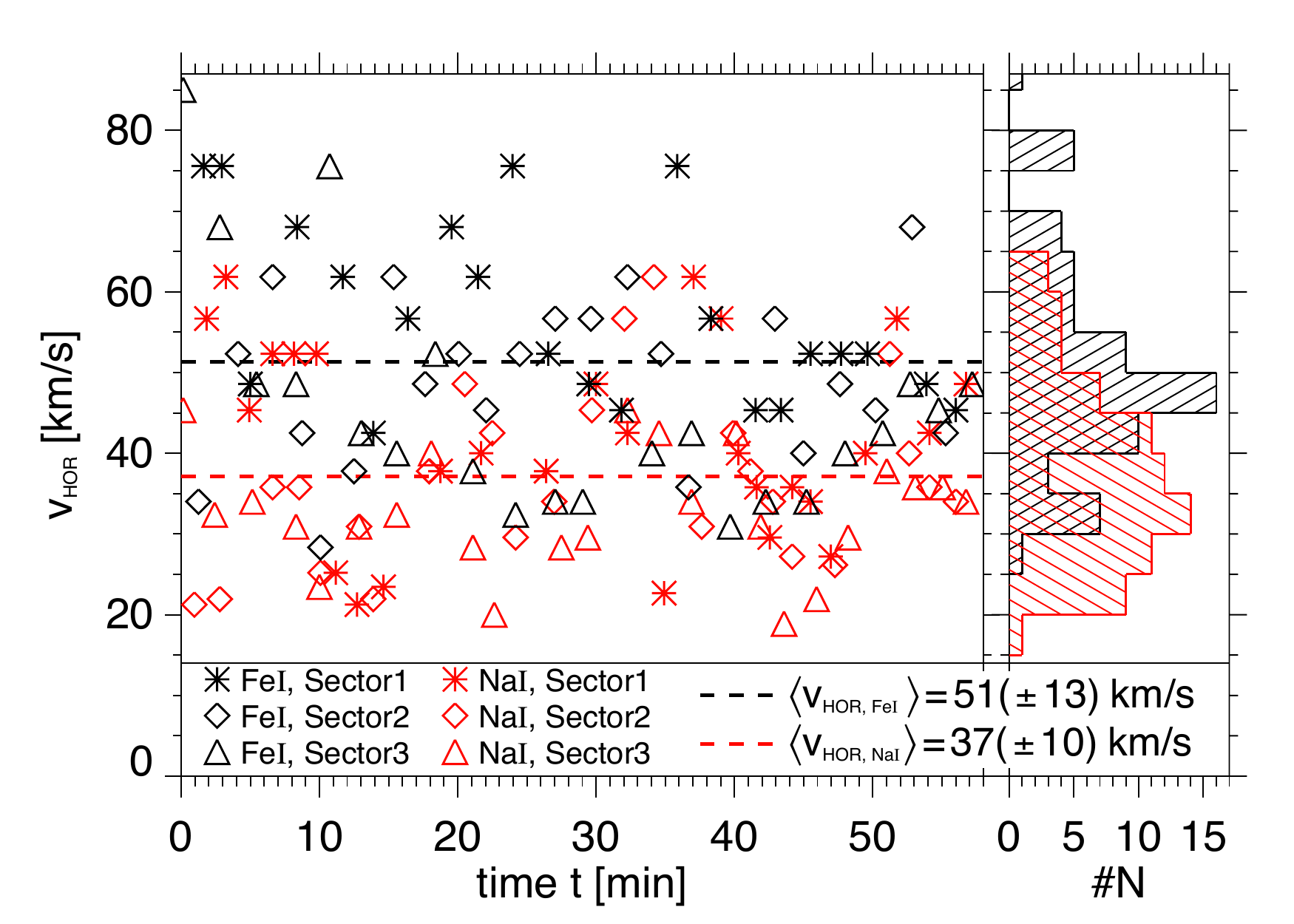}
\caption{Apparent horizontal velocities, $\rm{v}_{\rm HOR}$, of running penumbral waves at two atmospheric layers. The results for the photospheric \ion{Fe}{I} line (Fig.\,\ref{fig4}\,d-f) are shown in black, for the chromospheric \ion{Na}{I} line (Fig.\,\ref{fig4}\,a-c) in red. Left panel: the apparent horizontal velocities (in km/s) over the first $4\arcsec$ of the inner penumbra are plotted according to the observation time, $t$. The asterisks, diamonds, and triangles show the apparent speeds of the wave trains in sectors 1, 2, and 3. The dashed lines indicate the average apparent velocities, $\langle\rm{v}_{\rm HOR,\,\ion{Fe}{I}}\rangle$ and $\langle\rm{v}_{\rm HOR,\,\ion{Na}{I}}\rangle$ for both layers. The averages and their standard deviations are given in the figure legend. Right panel: histogram showing the number $\#N$ of velocity values $\rm{v}_{\rm HOR}$ from the left panel within a $5\,\rm{km/s}$ binning interval.}
\label{fig6}
\end{center}
\end{figure}

To prove the field-aligned propagation of RPWs in the lower sunspot atmosphere, we study the temporal evolution and trajectories of the waves. Therefore, we perform a sectorial time-slice analysis for the photospheric and chromospheric Doppler velocities, as shown in Fig.\,\ref{fig4}. To investigate the fluctuations, we subtract the temporally averaged Doppler velocities and focus on the relative velocities. Since the RPWs seem to follow the filamentary fine-structure of the penumbra and under the assumption of coherent wave trains for nearby fields, we select three circular (foreshortened) sectors with opening angles of $10^\circ$ for the penumbral regions with highest velocity power (see Fig.\,\ref{fig2}\,a,b). The temporal behavior of the azimuthal average at each distance from the spot center is shown in Fig.\,\ref{fig4} for the velocities derived from \ion{Na}{I} 589.6\,nm (panels a-c) and \ion{Fe}{I} 630.15\,nm (panels d-f). A detailed study of the oscillatory pattern demonstrates the photospheric origin of RPWs as follows:
\begin{itemize}
\item Continuous and clear signatures of penumbral oscillation at photospheric height are found.
\item Photospheric oscillations at the inner penumbra have an amplitude of up to $0.3\,\rm{km/s}$.
\item All sectors provide observational evidence for outward-directed horizontal propagation in the inner penumbra. 
\item In the inner penumbra close to the umbral boundary, all photospheric waves have a slightly delayed chromospheric counterpart with larger oscillatory amplitudes of up to $0.8\,\rm{km/s}$.
\item A simple analysis of the slopes of all wave trains over the first $4\arcsec$ of the inner penumbra was performed in the sectorial time slices shown in Fig.\,\ref{fig4}. The analysis reveals the apparent horizontal velocities, $\rm{v}_{\rm HOR}$, combined to Fig.\,\ref{fig6}. For the photospheric layers (black symbols and histogram), represented by the \ion{Fe}{I} 630.15\,nm line, the apparent horizontal velocities average to $\langle\rm{v}_{\rm HOR,\,\ion{Fe}{I}}\rangle=51\pm13\,\rm{km/s}$.
\item{At upper photospheric to chromospheric height, represented by the \ion{Na}{I} 589.6\,nm line (Fig.\,\ref{fig6}, red symbols and histogram), the wave trains indicate smaller apparent horizontal velocities of $\langle\rm{v}_{\rm HOR,\,\ion{Na}{I}}\rangle=37\pm10\,\rm{km/s}$. The smaller velocity is illustrated exemplarily in Fig.\,\ref{fig4} by the steeper slopes compared to the photospheric case.}
\end{itemize}

At chromospheric layers, the retrieved apparent horizontal velocities of around $20-50\,\rm{km/s}$ at the inner penumbra are in line with recent studies \citep[e.g., ][]{2006A&A...456..689T,2006SoPh..238..231K, 2013ApJ...779..168J}. Commonly,  this apparent propagation speed decreases radially toward the outer penumbra. As observations at higher chromospheric to transition region layers have shown \citep{2015ApJ...800..129M}, the retrieved horizontal velocities decrease to around $10\,\rm{km/s}$.\ Vice versa, the larger apparent velocities of around $30-70\,\rm{km/s}$ at photospheric layers fit well into the trend of decreasing horizontal velocities at the inner penumbra toward higher layers. The topological model of the sunspot's magnetic field \citep[e.g., ][]{1997Natur.389...47W}, the field-guided propagation of running penumbral waves, and their visual appearance at a certain layer can explain this behavior \citep{2007ApJ...671.1005B,2015ApJ...800..129M}. In the penumbra, the magnetic field inclination increases radially from the vertical umbral field. At photospheric layers, this bending of the magnetic field lines is stronger than at higher atmospheric layers in which the radial increase in field inclination happens more smoothly. As stated by \citet{2006RSPTA.364..313B}, the large apparent horizontal velocity of RPWs (here up to $80\,\rm{km/s}$ in Fig.\,\ref{fig6}) rather reflects relative travel time differences for waves guided by the individual less and more inclined magnetic field lines of the inner penumbra.

\section{Conclusions}\label{sec_conclusions}
We provide strong evidence for the presence of RPWs in the sunspot photosphere. Our observations conflict with the scenario that RPWs are a purely chromospheric wave phenomenon and strongly support the recent theory \citep[e.g.,]{2003A&A...403..277R,2007ApJ...671.1005B,2013ApJ...779..168J}
 that RPWs are upward propagating slow-mode waves guided by the magnetic field lines. Under the assumption of an expanding magnetic field topology with more inclined field lines with radial distance from the vertically oriented umbra, the performed wave power analysis substantiates the dependence of the power spectra on the atmospheric height and inclination of the magnetic field. 

The solar atmosphere exhibits a wealth of dynamical phenomena at all scales. These processes and their effects are strongly coupled. Especially in the case of sunspots, oscillations, flows, and other dynamical effects can  interact and have to be taken into account in observational studies \citep[e.g.,][]{2015ApJ...803...93E}.

A forthcoming, extensive study of the presented data (L\"ohner-B\"ottcher \& Bello Gonz\'alez, in preparation) will include an analysis of umbral flashes in the chromosphere (a bright umbral flash event can be seen in Fig.\,\ref{fig1}), their characteristic similarities and differences from RPWs, and the possible reconstruction of the magnetic field topology using the wave characteristics in sunspots. To verify the existence of photospheric RPWs, we suggest further observations with high spatial and temporal resolution using photospheric spectral lines that are magnetic insensitive (with a Land\'e-factor g=0), for example \ion{Fe}{I} 557.6\,nm \citep{2009A&A...508..941B} or \ion{Fe}{I} 543.4\,nm \citep{2010A&A...522A..31B}. For the latter, we have also found RPW signatures. As shown in Fig.\,\ref{fig3}, signatures of running penumbral waves can also be found in HMI Dopplergrams. For a quantitative evaluation of sunspot waves, we propose a statistical evaluation using a large sample of sunspots.
\begin{acknowledgements} The data were acquired in service mode operation within the transnational ACCESS program of SOLARNET, an EU-FP7 integrated activity project. The instruments IBIS and ROSA at the Dunn Solar Telescope (DST, NSO) were operated by INAF and QUB personnel, with special thanks to Gianna Cauzzi and Peter Keys. HMI data were used by courtesy of NASA/SDO and HMI science teams. This work was prepared at the Centre for Advanced Solar Spectro-polarimetric Data Analysis (CASSDA), funded by the Senatsausschuss of the Leibniz Association, Ref.-No. SAW-2012-KIS-5. We thank Wolfgang Schmidt for his fruitful comments on the manuscript.
\end{acknowledgements}
\bibliographystyle{aa} 
\bibliography{references}

\end{document}